\documentclass[prl,twocolumn,amsmath,amssymb,nofootinbib]{revtex4}
\usepackage{graphicx} 
\usepackage{epstopdf}
\usepackage{amsmath}
\usepackage{color}
\usepackage[caption=false]{subfig}
\usepackage{mathrsfs}
\usepackage{hyperref}

\usepackage{amsfonts}
\usepackage{amssymb}
\usepackage{bm}

\def\bea{\begin{eqnarray}}
\def\eea{\end{eqnarray}}
\def\be{\begin{equation}}
\def\ee{\end{equation}}
\def\<{\langle}
\def\>{\rangle}
\def\tr{\text{Tr}}
\def\dtp{{\delta_{\text{p}}}}
\def\ordr{\mathcal{O}}
\def\A{{\text{A}}}
\def\B{{\text{B}}}
\def\prob{{\mathcal{P}}}
\def\prbl{{\mathbf{f}}}

\def\non{\nonumber}
\def\sinc{{\text{sinc\,}}}

\def\el{{\text{el}}}
\def\inel{{\text{inel}}}

\begin{document}

\title{An Area Law for Entanglement Entropy in  Particle Scattering}

\author{\vspace{0cm} Ian Low$^{\, a,b}$, Zhewei Yin$^{\, b}$}
\affiliation{\vspace{0.1cm}
\mbox{$^a$ High Energy Physics Division, Argonne National Laboratory, Lemont, IL 60439, USA}\\
\mbox{$^b$ Department of Physics and Astronomy, Northwestern University, Evanston, IL 60208, USA}\\ 
 \vspace{-0.3cm}}

\begin{abstract}

The scattering cross section is the effective area of collision when two particles collide. Quantum mechanically, it is a measure of the probability for a specific process to take place.  Employing wave packets to describe the scattering process, we compute the entanglement entropy in 2-to-2 scattering of particles in a general setting using the $S$-matrix formalism. Applying the optical theorem, we show that the linear entropy $\mathcal{E}_2$ is given by  the elastic  cross section $\sigma_{\el}$ in unit of the transverse size $L^2$ of the wave packet, $\mathcal{E}_2 \sim \sigma_{\el}/L^2$, when the initial states are not entangled. The result allows for  dual interpretations of the entanglement entropy  as an area  and as a probability. Since $\sigma_{\el}$ is generally believed, and observed experimentally, to grow with the collision  energy $\sqrt{s}$ in the high energy regime, the result  suggests a ``second law'' of entanglement entropy for high energy collisions. Furthermore, the Froissart bound places an upper limit on the entropy growth.

\end{abstract}

\maketitle
\noindent


\section{Introduction}

Entropy is among the most fundamental concepts in physics and quantifies the number of possible microstates giving rise to a particular macrostate. It is directly related to probability -- a macrostate containing more microstates has a higher probability to occur and hence a higher entropy. A physical system naturally evolves into more probable states, thereby increasing the entropy, which  defines an arrow of time.

Entropy can also be understood as a measure of randomness, or disorder, since a more probable state is considered to be more disordered. As such the entropy is also used to quantify the amount of information in a physical system, since a highly disordered system is capable of carrying a large amount of information. Indeed, a sentence consists of a single alphabet is very ordered but conveys little information. In classical information theory the  Shannon entropy, which measures the average amount of information in an event drawn from a particular probability distribution \cite{Shannon:1948dpw,Nielsen:2012yss}, features  prominently.

One usually refers to the entropy defined in classical systems as coarse-grained entropy, which focuses on a subset of simple observables without knowing  the full microscopic configuration space. Even though the underlying dynamics is deterministic, the ignorance of the microstates necessitates the use of probability. Quantum mechanics, on the other hand, is inherently probabilistic and introduces intrinsic randomness in a quantum system. Such randomness is described by the quantum, or fine-grained, entropy \cite{Witten:2018zva,Almheiri:2020cfm}. The most prominent quantum entropy is the von Neumann entropy \cite{Nielsen:2012yss}, which can be generalized to two classes of quantum entropies: the R\'enyi entropy \cite{renyi} and the Tsallis entropy \cite{Tsallis}.

Similar to their classical counterpart, quantum entropies also quantify the amount of information present in a quantum system, as well as quantum correlations, or entanglement, between  systems \cite{Nielsen:2012yss}.  They are computed from the density matrix and also called entanglement entropies. While by definition the entropy of a closed system is invariant under unitary transformation, and hence time evolution, this is not always the case for an open quantum system interacting with an environment. In fact, one often divides a closed system into a bipartite system whose Hilbert space is ${\cal H}={\cal H}_\A\otimes {\cal H}_\B$. The subsystem entropy for  $\A$ is then computed from the reduced density matrix $\rho_\A$ by tracing over  $\B$, and can be nonzero even when the total  system has vanishing entropy.

In many-body systems the entropy is usually an extensive quantity and scales with the size of the system. Therefore it is quite mysterious that in gravitational systems, the black hole in particular, the  Bekenstein-Hawking entropy scales with the area of horizon \cite{Bekenstein:1973ur,Hawking:1975vcx}. Since then there have been  more examples of area laws for entropy both in gravitational and non-gravitational many-body systems \cite{Srednicki:1993im,Ryu:2006bv,Eisert:2008ur,Dvali:2017nis}, some of which  involves holography \cite{Maldacena:1997re,Almheiri:2020cfm}. In these examples the areas involved are often computed from a ``boundary'' between two ``regions.'' 

In this work we will consider the entanglement entropy in a very different setting: The 2-to-2 scattering of particles in quantum field theories, using the $S$-matrix formalism. This is hardly a many-body system, and it's not obvious at all if there is a ``boundary'' for which to define an area. Nevertheless, using the wave packet formalism to describe the particle scattering, we show that the linear entropy is given by the elastic cross section in unit of the transverse size of the wave packet when the initial states are not entangled. Recently the entanglement entropy in particle scattering has attracted considerable attention, although the calculation has mostly been done in specific quantum field theories and differing contexts \cite{Dvali:2014ila,Cervera-Lierta:2017tdt,Fan:2017mth,Beane:2018oxh,Tomaras:2019sjq,Dvali:2020wqi,Peschanski:2019yah,Aoude:2020mlg,Low:2021ufv,Dvali:2021ooc,Dvali:2021rlf,Muller:2022htn,Fedida:2022izl,Liu:2022grf,Hentschinski:2023izh,Cheung:2023hkq,Carena:2023vjc,Sakurai:2023nsc,Hu:2024hex,Aoude:2024xpx,Kowalska:2024kbs}. Here we will not commit to a specific  theory and will instead use the $S$-matrix formalism to make general statements which are non-perturbative in the couplings. Importantly, the wave packet formalism  gives a clear physical interpretation of the outcome.

Our result allows for novel, dual interpretations of the entanglement entropy as an area, in the cross-sectional sense, and a probability, in the information-theoretic sense. Moreover, it is generally accepted that the elastic cross section grows with the collision centre-of-mass (CM) energy $\sqrt{s}$  \cite{Cheng:1969tje,Cheng:1970bi}, which has been verified experimentally \cite{Block:1984ru,Pancheri:2016yel}. This suggests a version of the second law of thermodynamics where the entanglement entropy grows with energy, and  the Froissart bound \cite{Froissart:1961ux,Martin:1962rt} now gives an upper limit on the growth.

\section{The setup}

We consider the scattering of two distinguishable particles: A(lice) and B(ob). The kinematic data are labelled by the momentum $p_{\A/\B}$, which can be either massive or massless. All other non-kinematic quantum numbers are denoted by $\{ i \}$ for A and $\{\bar{i}\}$ for B, respectively. These could include spacetime quantum numbers, such as the spin, or the internal quantum numbers like the isospin.

The scattering process we focus on is $\A\B\to \A\B$, from which we construct a bipartite system ${\cal H}_{\A\B}={\cal H}_{\A}\otimes {\cal H}_{\B}$, where ${\cal H}_{\A/\B}$ is the Hilbert space associated with particle-$\A/\B$. For a general incoming state $|\rm{in}\rangle$, the density matrix is  $\rho^{\text{i}}=|\rm{in}\rangle\langle\rm{in}|$ and the  linear entropy ${\cal E}_2$ for  $\A$  is computed from the reduced density matrix $\rho_{\A}^{\text{i}} = {\rm Tr}_{\B} \ \rho^{\text{i}}$,
\be
{\cal E}_2=  1 - {\rm Tr}(\rho_\A^{\text{i}})^2 \ ,
\ee
which is the $n=2$ case of the Tsallis entropy ${\cal E}_n = (1-{\rm Tr}\,\rho^n)/(n-1)$ \cite{Tsallis}. From the von Neumann entropy ${\cal E}_{\text{vN}} = - {\rm Tr}(\rho \log \rho)$,  ${\cal E}_2$ can be obtained by expanding with respect to $\rho = 1$ and keeping the leading term. 

We will start with a pure initial state and comment on the mixed initial state later. We will also make the realistic assumption that there is no entanglement between  $p_\A$, $p_\B$, and between the momenta and the other quantum numbers. The initial state is then written as
\bea
|\text{in} \> = \sum_{i,\bar{i}} \Omega_{i \bar{i}} |\psi_{\A} \> \otimes | i\> \otimes |\psi_{\B} \> \otimes | \bar{i}\>,
\eea
where $|i \>$ and $|\bar{i} \>$ are normalized such that $\<i|j \> = \delta^{ij}$, $\<\bar{i} | \bar{j} \> = \delta^{\bar{i} \bar{j}}$, and $ \Omega$ describes the non-kinematic configuration of the initial state, which is normalized such that $\tr (\Omega \Omega^\dagger) = 1$. When the quantum numbers of two particles are not entangled, $\Omega_{i \bar{i}} = \omega_i \omega'_{\bar{i}}$ and $|\omega|^2 = |\omega'|^2= 1$. On the other hand, $|\psi_\A\>$ and $|\psi_\B\>$ are described by the momentum wave packets of $\A$ and $\B$, respectively,
\bea
| \psi_{\A/\B} \> = \int_{p} \psi_{\A/\B} (p)\, |p\>,\qquad \int_p \equiv \int \frac{d^3 \vec{p}}{(2 \pi)^3 \sqrt{2E_p}} \, ,
\eea
where the  one-particle state is normalized as follows,
\bea
\label{eq:deltap}
\<p| q\> = (2 \pi )^3\, 2 E_{p}\, \delta^3 (\vec{p} - \vec{q}).
\eea
The momentum wave packets are normalized such that
\bea
\label{eq:wavepac}
\< \psi | \psi \> = \int \frac{d^3 \vec{p}}{(2 \pi)^3} \ |\psi (p ) |^2 = 1.
\eea
With Eq.~(\ref{eq:wavepac}), the initial density matrix $\rho^{\text{i}}$ is  properly normalized, ${\rm Tr}\, \rho^{\text{i}} =1$. If we had used the one-particle  states $|{k}_\A\>$ and $|{k}_\B\>$ in place of $|\psi_\A\>$ and $|\psi_\B\>$, the $\delta$-function normalization in Eq.~(\ref{eq:deltap}) would have introduced divergences in ${\rm Tr}\, \rho$, which is often regularized by introducing a finite box in spacetime. We will not be using such a box normalization.  In the CM frame, $\vec{k}_\A = - \vec{k}_\B \equiv \vec{k}$. 

In an experimental setup the incoming particles are  prepared with  fairly well-defined momenta, subject to experimental resolution. Therefore we expect  $\psi(p)_{\A/\B}$ to be sharply peaked at $\pm \vec{k}$. That is, if $\dtp$ characterizes the linear width of the momentum wave packet,  we expect 
\be
{\dtp}/{|\vec{k}|} \ll 1 \ ,\qquad \psi ({k}) \sim 1/\dtp^{3/2} \ .
\ee
We compute the entanglement entropy in the limit $\dtp \to 0$ and to the first non-vanishing order in $\dtp/|\vec{k}|$.

Now let us consider the final state produced by the scattering process, which is given by $| \text{out} \> \equiv S |\text{in} \>$, where $S = 1 + iT$ is the scattering matrix and $T$ is the transition matrix, which is related to the scattering amplitude $M$:
\bea
 &&\< \{k_{\text{f}} \}, f_{\text{f}}| T | \{k_{\text{i}}\}, f_{\text{i}} \>\non\\
 &&= (2 \pi)^4 \delta^{4} \left(\sum k_{\text{f}} - \sum k_{\text{i}} \right) M_{f_{\text{i}}, f_{\text{f}}} (\{k_{\text{i}}\}; \{k_{\text{f}} \}),\label{eq:rtm}
\eea
where $f_{\text{i}/\text{f}}$ labels the discrete quantum numbers of the initial and final states. Unitarity of the $S$-matrix implies $2\, \text{Im}\, T = T^\dagger T$, which gives rise to the optical theorem.

In general, $|{\rm out}\>$ is a linear superposition of all possible outcome from the collision of $\A$ and $\B$. Since we will focus on the elastic scattering $\A\B\to \A\B$, the outgoing state is also consisted of  $\A$ and  $\B$, which can be selected  by manually applying a projection operator $P_{\A \B} $ on $|\text{out}\>$, $|{\rm out}\>_{\rm el} = P_{\A \B}|\text{out}\>$. The density matrix $\rho^{\text{f}}$ constructed  out of $|{\rm out}\>_{\rm el}$, however, is not properly normalized,
\bea 
\tr\left( P_{\A \B}|\text{out} \> \< \text{out}|P_{\A \B}\right) = \< \text{out} |P_{\A \B} | \text{out} \> = 1 - \mathcal{P}_{\inel}\, ,\ 
\eea
where
\bea
\prob_{\inel} = \< \text{out} |1-P_{\A \B} | \text{out} \>= \< \text{in}| T^\dagger (1 - P_{\A\B}) T |\text{in} \>\label{eq:c2d}
\eea
is the probability for $\A\B$ to scatter inelastically into  anything {\em but} $\A\B$. The properly normalized $\rho^{\text{f}}$ is now
\bea
\rho^{\text{f}} = \frac{1}{1-\prob_{\inel}}|\text{out} \>_{\rm el}\ _{\rm el} \< \text{out}|\ , \label{eq:fdmpn}
\eea
which we compute to leading order in $\dtp/|\vec{k}|$.

\section{Entanglement entropy}

The initial state reduced density matrix for A is
\bea
\rho^{\text{i}}_\A &=&  \sum_{i,j, \bar{i},\bar{j}} \Omega_{i\bar{i}} (\Omega^\dagger)_{\bar{j}j} \ \< \psi_B, \bar{j}| \psi_B, \bar{i} \>\  |\psi_A, i \>\<\psi_A, j|  \non\\
&=&\sum_{i,j} (\Omega  \Omega^\dagger)_{ij}  |\psi_A, i \>\<\psi_A, j| ,
\eea
and the entanglement entropy is
\bea
\mathcal{E}^{\text{i}}_2 = 1- \tr (\Omega^\dagger \Omega )^2.
\eea
Clearly, if there is no entanglement between the initial  states, $(\Omega^\dagger \Omega )^2 = \Omega^\dagger \Omega$ and the entropy $\mathcal{E}^{\text{i}}_2 = 0$.

\begin{figure}[tbp]
     \centering
     \subfloat[In momentum space]{\raisebox{6ex}{\scalebox{0.6}{\includegraphics{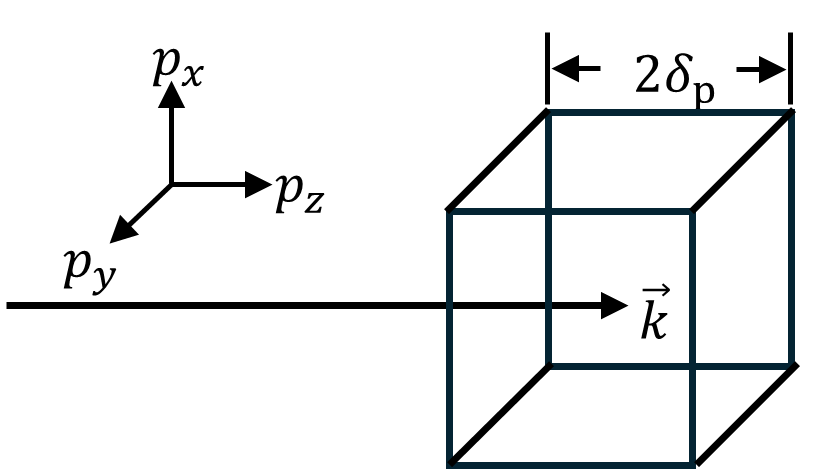}\label{fig:wpms}}}}
     \qquad
     \subfloat[In position space]{\scalebox{0.5}{\includegraphics{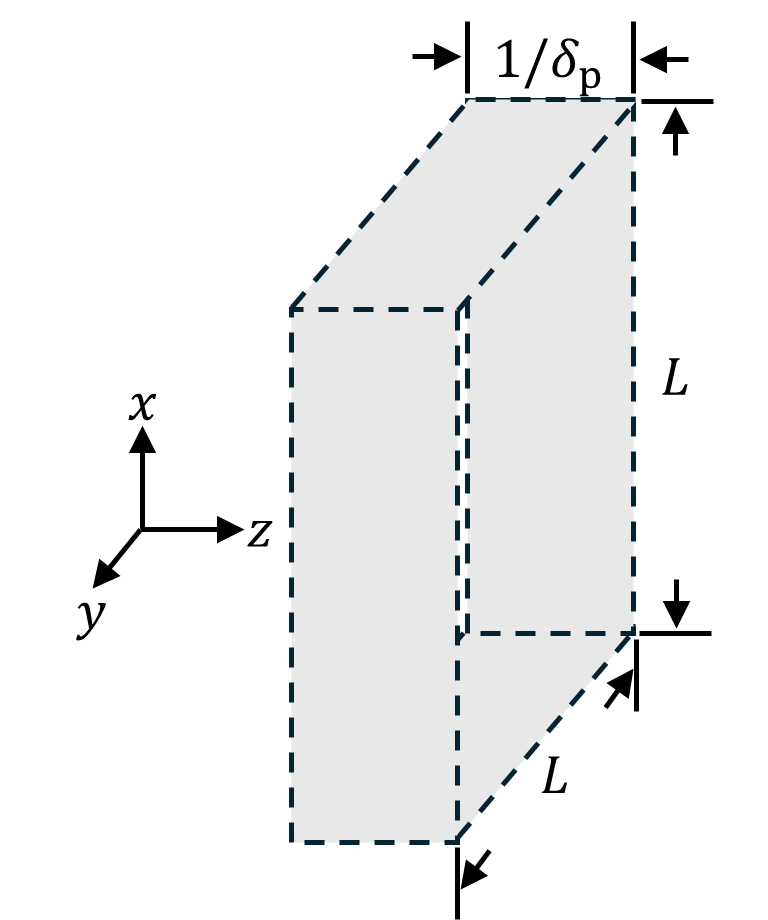}\label{fig:wpps}}}
        \caption{The wave packet in momentum space and in position space, respectively. In the limit $\dtp L\gg1$ and $\dtp\to 0$,  it approaches the plane wave in $z$-direction in the position space.
        }
        \label{fig:wp}
\end{figure}

To compute the final state reduced density matrix from Eq.~(\ref{eq:fdmpn}),  we choose the momentum-space  wave packet to be a cube with a side length of $2\dtp$, as shown in Fig. \ref{fig:wpms}. The wave packet is chosen to be uniform in the $p_z$-direction but not in the $p_{x,y}$-directions as we would like the wave packet to approach the plane wave  in position space. Specifically, we  define the regularized delta functions
\bea
\tilde{\delta}^3 (\vec{p} ) &=& \tilde{\delta}_0 (p_x)\tilde{\delta}_0 (p_y) \tilde{\delta}_z (p_z)\, , \label{eq:pfdef}\\
\tilde{\delta}_0 (k) &=& \frac{\pi}{2}\frac{\Theta ( k + \delta_{\text{p}}) -  \Theta ( k - \delta_{\text{p}})}{\text{Si} (\dtp L/2)} \frac{L}{2 \pi} \sinc \frac{kL}{2} ,\label{eq:pfind}\\
\tilde{\delta}_z (k )  &=&   \frac{\Theta ( k + \delta_{\text{p}}) -  \Theta ( k - \delta_{\text{p}})}{2 \delta_{\text{p}}}\, ,
\eea
where $\sinc (k) = (\sin k)/k$ is the Fourier transform of a uniform rectangular function and ${\rm Si}(x) = \int_0^x dy\ \sinc (y)$. The parameter $L$ characterizes the linear size of the wave packet in the position space in the $xy$-plane. We assume $\dtp L\gg 1$ such that  $\lim_{\dtp \to 0 } \tilde{\delta}^{(3)} (k ) = \delta^{(3)} (k)$. The correctly normalized wave function is
\bea
\label{eq:waveAB}
\psi_{\A/\B} (\vec{p}) = \frac{8\sqrt{\pi^5 \dtp}}{L}  \tilde{\delta}^3 ( \vec{p} - \vec{k}_{\A/\B} ).\label{eq:wpf}
\eea
The wave packet in the position space  is shown in Fig.~\ref{fig:wpps}, which is uniform inside a cross-sectional square of size $L^2$, confined within the shaded sides, up to $\ordr (1/(\dtp L))$ corrections; in the $z$-direction it is unbounded, with the width of the peak characterized by $1/\dtp$. It is useful to set $1/(\dtp L) \lesssim \dtp/|\vec{k}|$, so that taking the plane wave limit amounts to expanding around the single small parameter $\dtp/|\vec{k}|$. This requirement is equivalent to $L\lambda\gg 1/\dtp^2$, where $\lambda$ is the Compton wavelength for momentum $k$. For plane waves this is certainly satisfied.

With details provided in the supplementary material, one can show
\bea
\prob_{\inel}= I_0 (|\vec{k}|) \left[ \sigma_{\inel}+ \ordr (\dtp/|\vec{k}|) \right],\label{eq:pbabg}
\eea
where $\sigma_{\inel}$ is the inelastic cross section for  $\A\B\to$ anything but AB, and
\bea
\label{eq:i0}
I_0 (|\vec{k}|) &=& 4 |\vec{k}| \sqrt{s} \int_{p_1, p_2, q_1, q_2} \psi_\A (p_1) \psi_\B (p_2)\psi_\A^* (q_1)  \non\\
&&   \times \psi_\B^* (q_2)  (2 \pi)^4 \delta^{4} (q_1 + q_2 - p_1 - p_2) 
\eea
For our choice of wave packets  in Eq.~(\ref{eq:waveAB}), we show in the Supplementary Materials that 
\bea
I_0 (|\vec{k}|) = \frac{1}{L^2} \left( 1 + \mathcal{O} (\dtp/|\vec{k}|)\right).
\eea
In other words,  $I_0 (|\vec{k}|)$ is  the inverse of the cross-sectional area of the wave packet at  the leading order in $\dtp\to 0$,  in which limit $L\gg 1/\dtp$. The physical picture here is that the wave packet is traveling at a high speed in the $z$-direction, which is Lorentz-contracted, so in the CM frame the two wave packets look like two thin pancakes whose linear size in the transverse direction is much bigger than that in the longitudinal direction. In the strict plane wave limit, when $L\to \infty$, the two incoming wave packets are so spread out in the transverse direction in position space that the probability of them  interacting with each other is very small. This consideration suggests we can expand $\prob_{\inel}$    in Eq.~(\ref{eq:fdmpn})  as a small parameter.

Putting everything together, the final state entanglement entropy in the $\dtp\to 0$ limit is
\bea
\mathcal{E}^{\text{f}}_2 &=& 1- \tr (\Omega^\dagger \Omega )^2 (1 + 2\prob_{\inel})  \non\\
&&+\frac{I_0 (|\vec{k}|) }{|\vec{k}|\sqrt{s}} \left\{ \text{Im} \ \tr \left[ \Omega^\dagger \Omega \Omega^\dagger M^{\text{F}} (\Omega)   \right]  + \ordr (\dtp/|\vec{k}|) \right\} \non\\
&=& 1 - \tr (\Omega^\dagger \Omega )^2 + \frac{I_0 (|\vec{k}|) }{  |\vec{k}|\sqrt{s} } \bigl\{\text{Im} \ \tr \left[ \Omega^\dagger \Omega \Omega^\dagger M^{\text{F}} (\Omega)   \right]  \non\\
&&- 2|\vec{k}|\sqrt{s}\, \sigma_{\inel}\ \tr (\Omega^\dagger \Omega )^2 + \ordr (\dtp/|\vec{k}|)\bigr\}  ,\label{eq:eame}
\eea
where $[ M^{\text{F}} (\Omega)]_{i\bar{i}} =  \sum_{j,\bar{j}} \Omega_{j\bar{j}}   M_{j\bar{j},i\bar{i}}^\text{F}$ and
\bea
M_{i\bar{i},j\bar{j}}^{\text{F}} &=& M_{i\bar{i},j\bar{j}} (k_\A, k_\B; k_\A, k_\B)
\eea
is the elastic forward scattering amplitude   for  $\A(i) \B(\bar{i}) \to \A(j) \B(\bar{j})$. We demonstrate in the Supplementary Materials that terms dropped in Eq.~(\ref{eq:eame}) are higher orders in $\dtp/|\vec{k}|$. Now consider the case where the initial  states are not entangled; specifically, $\Omega_{i \bar{i}} = \omega_i \omega'_{\bar{i}}$, where
\bea
\omega = ( 1,0,0,\cdots ), \quad \omega' = ( 1,0,0,\cdots ).\label{eq:defc}
\eea
Then
\bea
\mathcal{E}^{\text{f}}_2 &=& I_0 (|\vec{k}|)  \frac{\text{Im} ( M^{\text{F}}_{1\bar{1},1\bar{1}} ) -  2|\vec{k}|\sqrt{s}\, \sigma_{\inel}+\ordr (\dtp/|\vec{k}|)}{ |\vec{k}|\sqrt{s} } \non\\
&=&2 I_0 (|\vec{k}|) \left[\sigma_{\text{tot}} -  \sigma_{\inel}  +\ordr (\dtp/|\vec{k}|) \right] \non\\
&=&2\frac{\sigma_{\el}}{L^2}   + \ordr (\dtp^5/ |\vec{k}|^5)\, 
,\label{eq:eadsw} 
\eea
where we have used the optical theorem to rewrite $\text{Im} ( M^{\text{F}}_{1\bar{1},1\bar{1}} )$ in terms of the total cross section $\sigma_{\text{tot}}$, and $\sigma_{\el}$ is now the elastic cross section for $\A\B\to \A\B$. It is interesting to observe that ${\sigma_{\el}}/{L^2}$ (at the leading order) is nothing but the probability for the two-to-two elastic scattering to occur,  
\bea
\frac{\sigma_{\el}}{L^2} + \ordr (\dtp^5/ |\vec{k}|^5) = \prob_{\el} = \< \text{in}| T^\dagger P_{\A\B} T |\text{in} \>,\ \ \label{eq:ptab}
\eea
which can be seen from the definition of cross section, 
$\sigma_{\el} = N_{\el}/(d_\A l_\A d_\B l_\B A)$, 
where $d_{\A/\B}$ are the number density of uniform beams, $l_{\A/\B}$ are the length of these beams, $A$ is the cross sectional area of both beams and $N_{\el}$ is the expectation value for the number of elastic scattering events. In our case we are scattering two single particles head-to-head, thus $d_\A l_\A A = d_\B l_\B A = 1$, $A = L^2$ and $N_{\el} = \prob_\el$.
The relation $\mathcal{E}_2 \sim \prob_\el$ is actually insensitive to the details of the wave packets, as long as we stay in the plane wave limit \cite{forthcoming}.

The factor of $2$ in Eq.~(\ref{eq:eadsw}) clearly comes from the fact that the linear entropy is computed from the second power of the density matrix. Then the computation can be directly generalized to e.g. the $n$-Tsallis entropy $\mathcal{E}_n(\rho) =(1- \tr\, \rho^n )/(n-1)$ where $n \ge 2$ is an integer:
\bea
 \mathcal{E}_n(\rho^{\text{f}}_\A) &=& \frac{n}{n-1}\frac{\sigma_{\el}}{L^2} + \ordr (\dtp^5/ |\vec{k}|^{5}).
\eea

We have been focusing on pure initial states; now we discuss briefly the case of mixed initial states. Let us consider unentangled states characterized by the following density matrix:
\bea
\rho^{\text{i}} = \rho^{\text{i}}_{f_\A} \otimes \rho^{\text{i}}_{p_\A} \otimes \rho^{\text{i}}_{f_\B} \otimes \rho^{\text{i}}_{p_\B},
\eea
where $\rho^{\text{i}}_{f_{\A/\B}}$ and $\rho^{\text{i}}_{p_{\A/\B}}$ are density matrices for the discrete and momentum subsystems of the two initial particles, respectively. Let us consider mixed  states in discrete quantum numbers, such as the spin:
\bea
\rho^{\text{i}}_{f_\A} = \sum_{i} \prbl_{i}\ | i \> \< i |, \qquad \rho^{\text{i}}_{f_\B} = \sum_{\bar{i}} \bar{\prbl}_{\bar{i}}\ | \bar{i} \> \< \bar{i} |,
\eea
where  $\prbl_{i}\ge 0$, $\bar{\prbl}_{\bar{i}} \ge 0$, and $\sum_i \prbl_{i} = \sum_{\bar{i}} \bar{\prbl}_{\bar{i}} = 1$. Suppose the momentum states are still pure: $\rho^{\text{i}}_{p_\A} = |\psi_\A \> \< \psi_\A|$, $\rho^{\text{i}}_{p_\B} = |\psi_\B \> \< \psi_\B|$, the subsystem entropy of A is 
\bea
\mathcal{E}^{\text{f}}_{2,\A} = 2 I_0 (|\vec{k}|) \sum_{i,\bar{i}}\, \prbl_{i}^2\, \bar{\prbl}_{\bar{i}}\, \left[ (\sigma_{\el})_{i\bar{i}}+\ordr (\dtp/|\vec{k}|)\right]  .
\eea
where $(\sigma_{\el})_{i\bar{i}}$ is the total elastic cross section where the initial discrete quantum numbers are $i$ and $\bar{i}$. On the the hand, the  elastic scattering probability for colliding mixed states is
\bea
\prob_{\el} &=&\frac{1}{L^2} \sum_{i,\bar{i}} \prbl_{i} \bar{\prbl}_{\bar{i}} (\sigma_{\el})_{i\bar{i}} +\ordr (\dtp^5/|\vec{k}|^5) \non\\
&\equiv&  \frac{\overline{\sigma_{\el}}}{L^2} + \ordr (\dtp^5/ |\vec{k}|^5)\, ,
\eea
from which we see that, when the density matrix satisfies $(\rho^{\text{i}}_{f_\A})^2 \propto \rho^{\text{i}}_{f_\A}$, the  subsystem entropy can again be interpreted as a cross section,
\bea
\mathcal{E}^{\text{f}}_{2,\A} = \frac{2}{n_\A} \frac{\overline{\sigma_{\el}}}{L^2} + \ordr (\dtp^5/ |\vec{k}|^5)  .
\eea
The condition $(\rho^{\text{i}}_{f_\A})^2 \propto \rho^{\text{i}}_{f_\A}$ is satisfied  when $\prbl_{i}$ is  zero or the non-zero entries are all given by  the same constant value, $\prbl_{i} = 1/n_\A$, where $n_\A$ counts the number of nonzero entries in $\prbl_{i}$.
The same criteria was proposed in Ref.~\cite{Cheung:2023hkq}, which  ensures the subsystem entropy to not decrease upon time evolution. A  realistic scenario fulfilling this criteria is the unpolarized scattering, in which case $\overline{\sigma_{\el}}$ is simply the unpolarized elastic cross section. 

\section{Discussions and Conclusion}

Having shown the entanglement entropy in 2-to-2 particle scattering is given by the elastic cross section in units of the transverse size of the wave packets, several interesting observations follow.

Since the cross section is the effective area of collision when particles collide, we have essentially discovered an area law for the entanglement entropy in a two-body system. In other examples of area laws for the entropy, the area is often given by  the boundary of a spatial region \cite{Bekenstein:1973ur,Hawking:1975vcx,Srednicki:1993im,Ryu:2006bv,Eisert:2008ur,Dvali:2017nis}. In the present case it is not clear if a similar interpretation could be given. We speculate that, perhaps, the cross section could be viewed as the area of the space-like surface connecting the past and the future light cones of the two particle scattering. On the other hand, cross section measures the probability that a quantum mechanical process would happen, in accordance with the thermodynamic and information-theoretic interpretations of the entropy as probability.

Scaling properties of cross sections, in particular the total and elastic cross sections, with respect to the collision energy $\sqrt{s}$,  have been studied intensively for high energy collisions many decades ago. Froissart and Martin showed that the total cross section is bounded from the above by $\log^2 s$ \cite{Froissart:1961ux,Martin:1962rt} in the $S$-matrix formalism. Cheng and Wu argued very generally in quantum field theory that both the total and the elastic cross sections rise with $\sqrt{s}$ \cite{Cheng:1969tje,Cheng:1970bi}, which is a somewhat surprising result because cross sections for elementary processes at a fixed order  in perturbation always decrease with  $\sqrt{s}$. However, in a high energy scattering process there is abundant energy to create radiations and splittings of soft and collinear particles, which necessitates the resummation of higher order diagrams, akin to the soft photon resummation in quantum electrodynamics. These effects enhance the cross section to increase with the energy in the form of $\log s$. In fact, Ref.~\cite{Cheng:1970bi} proposed a ``Black Disk'' picture of a total absorption area of radius $\log s$ for high energy collisions, which is similar in spirit to the wave packet adopted in this work. The growth  of the cross section with respect to $\sqrt{s}$ has been measured over several orders of magnitude in $\sqrt{s}$ in $pp$ and $p\bar{p}$ collisions at the Tevatron and the Large Hadron Collider \cite{Block:1984ru,Pancheri:2016yel}. (See, for example, Fig.~5.40 in Ref.~\cite{Pancheri:2016yel}.) Even though many of the studies are based on hadron-hadron collisions, it has become clear recently in the study of a multi-TeV muon collider that radiations and splittings are also important in high energy collisions of leptons \cite{Chen:2016wkt,Han:2020uid,Ruiz:2021tdt}. Therefore, similar considerations for cross sections for lepton-lepton collisions should apply.

These observations lead to a  ``second law'' of entanglement entropy  with respect to the CM energy $\sqrt{s}$, in which the area (cross section) grows with the energy and so does the entanglement entropy. The Froissart bound now constrains the growth of the  entropy with respect to energy. It would be interesting to explore whether there is a holographic understanding of the relation between the entanglement entropy and the cross section. 

Last but not least, certain other cross sections, for instance the Deep Inelastic Scattering-like cross section, can also be related to entanglement entropies corresponding to different ways to construct a bipartite system in 2-to-2 scatterings \cite{forthcoming}. It will be interesting to see if all cross sections can be interpreted this way.

\section{Acknowledgement}

Discussions with Tao Han on the scaling properties of cross sections are gratefully acknowledged. We also thank Gia Dvali and Minglei Xiao for comments on the manuscript. This work  is supported in part by the U.S. Department of Energy, Office of High Energy Physics, under contract DE-AC02-06CH11357 at Argonne, as well as by the U.S. Department of Energy, Office of Nuclear Physics, under grant DE-SC0023522 at Northwestern.

\bibliography{references}

\clearpage

\begin{appendix}

\section{Supplementary Materials: Wave Packet Treatment}

\label{app:wpc}

Here we detail  the computations involving wave packets. First, let us see that in general, how the universal factor $I_0 (|\vec{k}|)$ appears at the leading order in the relevant computations. For example, consider $\prob_{\inel}$ in Eq.~(\ref{eq:c2d}), which is given by
\bea
{\prob}_{\inel} &=&   \int_{p_1, p_2, q_1, q_2}   \psi_\A (p_1) \psi_\B (p_2) \psi_\A^* (q_1) \psi_\B^* (q_2) \non\\
&&\!\!\!\!\!\!\!\! \!\!\!\!\!\!\!\!\!\!\!\!\!\!\!\!\!\!\!\! \times (2\pi)^4 \delta^4 (q_1 + q_2 - p_1 - p_2)\, {\cal F}_{\inel} (p_1, p_2, q_1, q_2)\, ,\,\label{eq:wfgep}
\eea
with
\bea
&&(2\pi)^4 \delta^4 (q_1 + q_2 - p_1 - p_2)\, {\cal F}_{\inel} (p_1, p_2, q_1, q_2)\non\\
&=&\sum_{i,\bar{i},j,\bar{j}} \Omega_{i \bar{i}} \Omega^*_{j \bar{j}}\<q_1,j;q_2,\bar{j}|T^\dagger (1 - P_{\A \B}) T|p_1,i;p_2,\bar{i}\>,\quad\,\,\,
\eea
which can be expressed in terms of inelastic amplitudes using Eq. (\ref{eq:rtm}). Now, without specifying the detail of the wave packets, we require that in momentum space the wave functions are bounded within an $\ordr (\dtp)$ radius, thus
\bea
{\cal F}_{\inel} (p_1, p_2, q_1, q_2) &=& {\cal F}_{\inel} (k_\A, k_\B, k_\A, k_\B) \non\\  
     && \qquad \times [1 + \mathcal{O} (\dtp/|\vec{k}|) ]\,
\eea
can be pulled out of the integral in Eq. (\ref{eq:wfgep}):
\bea
&&{\prob}_{\inel} =   {\cal F}_{\inel} (k_\A, k_\B, k_\A, k_\B) \left[1 + \mathcal{O} (\dtp/|\vec{k}|) \right] \non\\
&&\times\int_{p_1, p_2, q_1, q_2}   \psi_\A (p_1) \psi_\B (p_2) \psi_\A^* (q_1)  \psi_\B^* (q_2) \non\\
&&\times (2\pi)^4 \delta^4 (q_1 + q_2 - p_1 - p_2)\non\\
&&= I_0 (|\vec{k}|)  \left[ \sigma_{\inel} +\ordr (\dtp/|\vec{k}|) \right],
\eea
and we arrive at  Eq.~(\ref{eq:pbabg}).  In general, we see in computations other wave packet integrals similar to $I_0 (|\vec{k}|)$. However, we will argue that all other  integrals encountered are higher in $\dtp$ and can be neglected.

Now let us consider the case of uniform wave functions in the transverse directions of the position space. Specifically, we define $\tilde{\delta}^3 (\vec{p} )$ as in Eq.~(\ref{eq:pfdef}) and the wave packet as in Eq.~(\ref{eq:waveAB}). The wave function is chosen in the following spirit: We first bound the wave function strictly inside a cube of size $ (2\dtp)^3$ in the momentum space, as shown in Fig. \ref{fig:wpms}. Next, we treat the transverse directions separately compared to the $z$ direction, as in Fig. \ref{fig:wpps}: We make the wave function in the $z$ direction uniform within the $2\dtp$ sized window in the momentum space. On the other hand, we require that the wave function is bounded and uniform within a square of size $L^2$ in the $x$- and $y$-direction of the position space, up to $\ordr (1/(\dtp L))$ corrections; we choose $L \gg 1/\dtp$, with $1/\dtp$ characterizing the width of the peak along the $z$ direction in the position space. We also set $1/(\dtp L) \lesssim \dtp/|\vec{k}|$ so  there is a single small parameter $\dtp/|\vec{k}|$ to expand.

Let us first consider the general integral below:
\bea
\label{eq:i0tilde}
&&\!\!\!\!\!\!\tilde{I}_0 = \int d^3 \vec{p}_1 d^3 \vec{p}_2 d^3 \vec{q}_1 d^3 \vec{q}_2 \ \delta^{4} (q_1 + q_2 - p_1 - p_2)\non\\
&&\!\!\!\!\!\!\times {\cal F}_0 (p_1, p_2, q_1, q_2)\non\\
&&\!\!\!\!\!\!\times \tilde{\delta}^3 (\vec{p}_1 - \vec{k}) \tilde{\delta}^3 (\vec{p}_2 + \vec{k}) \tilde{\delta}^3 (\vec{q}_1 - \vec{k}) \tilde{\delta}^3 (\vec{q}_2 + \vec{k}) . \quad
\eea
As previously argued, we can pull ${\cal F}_0$ out of the integral at the leading order (assuming ${\cal F}_0 (k_\A, k_\B, k_\A, k_\B)$ to be non-vanishing):
\bea
&&\!\!\!\!\!\!\tilde{I}_0 =  {\cal F}_0 (k_\A, k_\B, k_\A, k_\B)\left[1 + \mathcal{O} (\dtp/|\vec{k}|) \right] \non\\
&&\!\!\!\!\!\!\times\int d^3 \vec{p}_1\, d^3 \vec{p}_2\, d^3 \vec{q}_1\, d^3 \vec{q}_2 \   \tilde{\delta}^3 (\vec{p}_1 - \vec{k})   \tilde{\delta}^3 (\vec{p}_2 + \vec{k}) \non\\
&&\!\!\!\!\!\!\times \tilde{\delta}^3 (\vec{q}_1 - \vec{k}) \tilde{\delta}^3 (\vec{q}_2 + \vec{k}) \delta^{4} (q_1 + q_2 - p_1 - p_2) .
\eea
We can first integrate out the spatial $\delta$-functions along the $\vec{p}_1$ direction, and then integrate out the energy $\delta$-function in the $(p_2)_z$ direction:
\bea
&&\delta (E_{p_1} + E_{p_2} - E_{p_3} - E_{p_4}) \non\\
&&\to \left( \frac{E_{k_\A} E_{k_\B}}{ |\vec{k}| \sqrt{s} } + \mathcal{O} (\dtp/|\vec{k}|) \right) \delta ( (p_2)_z - r_0),
\eea
where the factor in front of the $\delta$-function on the right hand side of the above can be pulled out of the integral like ${\cal F}_0$, and $r_0$ is the root of $(p_2)_z$ in
\bea
&&\sqrt{(\vec{q_1} + \vec{q_2} - \vec{p_2})^2 + m_\A^2 } + \sqrt{\vec{p}_2^2 + m_\B^2}  \non\\
&&\qquad\qquad - \sqrt{\vec{q}_1^2 + m_\A^2}- \sqrt{\vec{q}_2^2 + m_\B^2} = 0.\label{eq:edfrz}
\eea
Then the integration we need in the $(p_2)_z$ direction is 
\bea
&&\int d (p_2)_z \delta ( (p_2)_z - r_0) \tilde{\delta}_z ( (p_2)_z + | \vec{k}|) \non\\
&=& \frac{\Theta (r_0 + | \vec{k}| + \dtp) - \Theta (r_0 + | \vec{k}| - \dtp)}{2\dtp} ,
\eea
which is non-vanishing within a $2 \delta_p$ sized window if we vary $r_0$ in the above. Therefore, to get leading order effects in $\dtp$, we need to solve $r_0$ up to linear order in $\dtp$, which turns out to be $r_0 = (q_2)_z + \mathcal{O} (\dtp^2/|\vec{k}|^2)$.\footnote{One can consider characterizing the size of the wave packet in the $p_z$ direction by some $\delta_z$, which is independent of $\dtp$. Then the solution $r_0$ needs to be computed up to $\ordr (\delta_z/|\vec{k}|)$, but it will admit an expansion of $\max (\dtp, \delta_z)$. Therefore, we need to require $\delta_z \gtrsim \dtp$ for our computation to still hold. In other words, $1/\dtp $ is the largest possible length scale which characterizes the spread of the wave packet in the direction of the momentum. As $1/\dtp \ll L$, the wave packet is much more compressed in the $z$ direction compared to the transverse directions.} Then one sees that the integration of the different spatial dimensions actually decouples.

In the transverse directions, notice that if we take $1/(\dtp L) \to 0$ in Eq.~(\ref{eq:pfind}), we have
\bea
\int dk \ \tilde{\delta}_0 (k) \to \int_{-\infty}^{\infty} d (kL) \frac{1}{2 \pi}\, \sinc \frac{kL}{2}.
\eea
Therefore, as we are assuming $\dtp \gg 1/L$, at the leading order   the $\Theta$-functions as well as the normalization factor related to $\text{Si}(x)$ in Eq.~(\ref{eq:pfind}) can be effectively omitted from our computation:
\bea
\tilde{\delta}_0 (k) &=& \frac{\pi}{2}\frac{\Theta ( k + \delta_{\text{p}}) -  \Theta ( k - \delta_{\text{p}})}{\text{Si} (\dtp L/2)} \frac{L}{2 \pi} \,\sinc \frac{kL}{2}\non\\
&\sim& \frac{L}{2 \pi}\, \sinc \frac{kL}{2}\ .
\eea
The above approximation is valid up to $\ordr (1/(\dtp L))$, and as we would like an expansion of small $\dtp/|\vec{k}|$, it is then desirable to have $1/(\dtp L) \lesssim \dtp/|\vec{k}|$. The useful integral in the transverse directions is then
\bea
&&\int d (p_1)_x \int d (p_2)_x \int d (q_1)_x \int d (q_2)_x \non\\
&& \times \delta \left[( \vec{p}_1+ \vec{p}_2- \vec{q}_1 - \vec{q}_2)_x \right]  \sinc \frac{L(p_1)_x}{2} \sinc \frac{L(p_2)_x}{2} \non\\
&& \times\sinc \frac{L(q_1)_x}{2} \sinc \frac{L(q_2)_x}{2} = \left(\frac{2 \pi}{L} \right)^3 .
\eea
In the $z$ direction we need
\bea
 &&\int_{k-\dtp}^{k+\dtp} d (p_1)_z \int_{-k-\dtp}^{k+\dtp} d (p_2)_z  \int_{k-\dtp}^{k+\dtp} d (q_1)_z \non\\
 &&\times\int_{-k-\dtp}^{-k+\dtp} d (q_2)_z  \delta \left[( \vec{p}_1+ \vec{p}_2- \vec{q}_1 - \vec{q}_2)_z \right] \delta [(\vec{p}_2 - \vec{q}_2)_z]\non\\
 &&\ = 4 \dtp^2 .
\eea
Then
\bea
\tilde{I}_0 &=&  \frac{1 }{16 \pi^2}  \frac{ E_{k_\A} E_{k_\B}}{ |\vec{k}| \sqrt{s} } \frac{L^2}{\dtp^2} \non\\
&&\times{\cal F}_0 (k_\A, k_\B, k_\A, k_\B)\left[ 1 + \mathcal{O} (\dtp/|\vec{k}|) \right].
\eea

Now we can compute $I_0 (|\vec{k}|)$ from $\tilde{I}_0$. Let's plug in the explicit form of the wave packet from Eq.~(\ref{eq:waveAB}) into $I_0 (|\vec{k}|)$ in Eq.~(\ref{eq:i0}), and compare with $\tilde{I}_0$ in Eq.~(\ref{eq:i0tilde}), we see $I_0 (|\vec{k}|)$ is given by 
\bea
{\cal F}_0(k_\A, k_\B, k_\A, k_\B) \to \frac{2^{12} \pi^{10} \dtp^2 4 |\vec{k} | \sqrt{s}  }{(2 \pi)^{8} 2E_{k_\A} 2E_{k_\B} L^4 } .
\eea
In the end we arrive at
\bea
I_0 (|\vec{k}|) = \frac{1}{L^2} \left( 1 + \mathcal{O} (\dtp/|\vec{k}|)\right).
\eea
We see clearly here that at the leading order, $I_0 (|\vec{k}|)$ is exactly the inverse of the cross-sectional area of the wave packet.

Now let us see concretely  other types integrals involving the overlap of wave packets which appear in the entanglement entropy calculation. We will show that they are higher order in $\dtp/|\vec{k}|$ and can be neglected. For example, $\mathcal{E}^{\text{f}}_{2}$ in Eq. (\ref{eq:eame}) contains the following terms:
\bea
2 \, \text{Re}\ \text{tr} \ ( \Omega^\dagger \tilde{M} (\Omega)  )^2  - 2\, \text{\tr} \left[ \Omega^\dagger \Omega \tilde{M}^{(2)}_{1} (\Omega) \right],\label{eq:hoe}
\eea
where
\bea
&&\left[\tilde{M} (\Omega) \right]_{i \bar{i}} = \sum_{j,\bar{j}} \Omega_{j\bar{j}}  \int_{p_1, p_2, q_1, q_2} \psi_\A (p_1) \psi_\B (p_2)\non\\
&& \times  \psi_\A^* (q_1) \psi_\B^* (q_2)  (2 \pi)^4 \delta^{4} (q_1 + q_2 - p_1 - p_2) \non\\
&&\times M_{j\bar{j},i\bar{i}} (p_1, p_2; q_1, q_2)  \non\\
&&= \frac{1}{ 4 |\vec{k}|\sqrt{s} {L^2}}  \Big[\sum_{j,\bar{j}} \Omega_{j\bar{j}}   M_{j\bar{j},i\bar{i}}^\text{F} + \mathcal{O} (\dtp/|\vec{k}|) \Big],\\
&&\left[\tilde{M}^{(2)}_{1} (\Omega) \right]_{\bar{i}_1 \bar{i}_2} =  \sum_{i,j, \bar{j}, k, \bar{k}} \int_{p_1, p_2, p_3, p_4,q_1,q_2,q_4} \frac{1}{\sqrt{2E_{p_3}}}\non\\
&&\Omega^*_{j \bar{j}} \Omega_{k \bar{k}} \psi^*_\A (p_1) \psi^*_\B (p_2)  \psi_\B (p_4) \psi_\A (q_1) \psi_\B (q_2) \psi^*_\B (q_4)\non\\
&& \times(2 \pi)^8 \delta^{4} (p_3 + p_4 - p_1 - p_2)  \delta^{4} (q_3 + q_4 - q_1 - q_2)   \non\\
&& \times     M^*_{j \bar{j},i \bar{i}_1} (p_1, p_2; p_3, p_4) \non\\
&&\times \left.M_{k \bar{k},i \bar{i}_2} (q_1, q_2; q_3, q_4)    \right|_{\vec{q}_3 = \vec{q}_1 + \vec{q}_2 - \vec{q}_4} .
\eea
Then the first term in Eq.~(\ref{eq:hoe})  is $\ordr (1/{(|\vec{k}|L)^4})\alt \ordr (\dtp^8/|\vec{k}|^{8})$ and can be dropped. (Recall we demand $1/(\dtp L) \lesssim \dtp/|\vec{k}|$.) The second term in Eq.~(\ref{eq:hoe}) involves the following integral:
\bea
&&\!\!\!\!\!\!\!\!\!\!\!\!\tilde{I}_1 = \int d^3 \vec{p}_1 d^3 \vec{p}_2 d^3 \vec{p}_3 d^3 \vec{p}_4 d^3 \vec{q}_1 d^3 \vec{q}_2  d^3 \vec{q}_4 \non\\
&&\!\!\!\!\!\!\!\!\!\!\!\!\times \delta^{4} (q_1 + q_2 - p_3 - q_4)\delta^{4} (p_1 + p_2 - p_3 - p_4) \non\\
&&\!\!\!\!\!\!\!\!\!\!\!\!\times {\cal F}_1 (p_1, p_2,p_3,p_4, q_1, q_2,q_4)\tilde{\delta}^3 (\vec{p}_1 - \vec{k}) \tilde{\delta}^3 (\vec{p}_2 + \vec{k}) \non\\
&&\!\!\!\!\!\!\!\!\!\!\!\!\times  \tilde{\delta}^3 (\vec{p}_4 + \vec{k})  \tilde{\delta}^3 (\vec{q}_1 - \vec{k}) \tilde{\delta}^3 (\vec{q}_2 + \vec{k}) \tilde{\delta}^3 (\vec{q}_4 + \vec{k}) .\ 
\eea
Here, $p_3$ is constrained by the on-shell condition of type-A particles, and the integrand in the above will constrain it into a range of radius $\ordr (\dtp)$ around $k_\A$. Similar to $\tilde{I}_0$, we pull out ${\cal F}_1$ first. Then we integrate out $\delta^3 (\vec{p_1} + \vec{p}_2 - \vec{p}_3 - \vec{p_4} )$ in the direction of $\vec{p}_3$, and the other spatial $\delta$-function in the direction of $\vec{q}_1$. Next, integrating out $\delta (E_{q_1} + E_{q_2} - E_{p_3} - E_{q_4})$ in the $(q_2)_z$ direction, we convert it to $\delta [(\vec{q}_2 - \vec{q}_4)_z]$; integrating out $\delta (E_{p_1} + E_{p_2} - E_{p_3} - E_{p_4})$ in the $(p_4)_z$ direction, we convert it to $\delta [(\vec{p}_4 - \vec{p}_2)_z]$. Again, the integrals in the spatial dimensions decouple, and upon integration we have
\bea
\tilde{I}_1 &=&   {\cal F}_1 (k_\A, k_\B, k_\A, k_\B,k_\A, k_\B, k_\B)\left[1 + \mathcal{O} (\dtp/|\vec{k}|) \right] \non\\
&&\times \frac{1}{32 \pi^2}\left(\frac{ E_{k_\A} E_{k_\B}}{ |\vec{k}| \sqrt{s} } \right)^2 \frac{L^2}{\dtp^3}. 
\eea
Therefore,
\bea
&&\left[\tilde{M}^{(2)}_{1} (\Omega) \right]_{\bar{i}_1 \bar{i}_2} =  \left(\frac{1}{ 4 |\vec{k}|\sqrt{s} {L^2}} \right)^2   \non\\
&& \times \Big[\sum_{i,j, \bar{j}, k, \bar{k}}  \Omega^*_{j \bar{j}} \Omega_{k \bar{k} } M^{\text{F}*}_{j \bar{j},i \bar{i}_1}  M^{\text{F}}_{k \bar{k},i \bar{i}_2} + \mathcal{O} (\dtp/|\vec{k}|) \Big],\ \ 
\eea
thus the second term in Eq. (\ref{eq:hoe}) is also {$\ordr (1/ {(|\vec{k}|L)^4})$} and can be dropped. Notice that we are dropping these higher order terms because we are taking the plane wave limit $\dtp\to 0$, instead of any weak coupling limit as in e.g. Ref.~\cite{Aoude:2024xpx}. Therefore, our computation is valid to all orders in the coupling strength.

\end{appendix}

\end{document}